# The Ising transition in 2D simplicial quantum gravity - can Regge calculus be right?[*]

Christian Holm[a] and Wolfhard Janke[a][b][†]

[a]Institut für Theoretische Physik, Freie Universität Berlin, 14195 Berlin, Germany

[b]Institut für Physik, Johannes Gutenberg-Universität Mainz, 55099 Mainz, Germany

We report a high statistics simulation of Ising spins coupled to 2D quantum gravity in the Regge calculus approach using triangulated tori with up to $512^2$ vertices. For the constant area ensemble and the $dl/l$ functional measure we definitively can exclude the critical exponents of the Ising phase transition as predicted for dynamically triangulated surfaces. We rather find clear evidence that the critical exponents agree with the Onsager values for static regular lattices, independent of the coupling strength of an $R^2$ interaction term. For exploratory simulations using the lattice version of the Misner measure the situation is less clear.

## 1. INTRODUCTION

Ising spins coupled to two-dimensional (2D) Euclidean quantum gravity in the dynamically triangulated surface (DTS) formulation exhibit a third-order phase transition with critical exponents which are very different from the Onsager values for static 2D lattices. This follows from exact results for matrix models [1] and numerical simulations [2], and is in perfect agreement with predictions from conformal field theory for matter of central charge $c = 1/2$ in the continuum limit [3].

An alternative formulation is the Regge calculus [4] where the connectivity of the discretized surface is fixed but the link lengths vary. Recent Monte Carlo (MC) simulations [5] of pure 2D gravity were interpreted in accordance with the DTS results. When coupled to Ising spins, however, evidence for Onsager critical exponents was claimed [5]. Since this would be a severe failure of the Regge approach we decided to investigate this question again with much higher precision [6]. Finally it should be remarked that already the consistency of pure gravity with the $dl/l$ measure was recently questioned in Ref.[7].

[*]Work supported in part by the EEC under contract No. ERBCHRX CT93043 and by the NVV computer grant bvpf01.
[†]WJ thanks the DFG for a Heisenberg fellowship.

## 2. MODEL AND SIMULATION

Employing the usual transcription [8] of continuum quantities like the metric $g$ and the scalar curvature $R$, the Regge partition function reads

$$Z = \sum_{\{s\}} \int \mathcal{D}\mu(l) \exp\left(-I(l) - KE(l,s)\right), \qquad (1)$$

where

$$I(l) = \sum_i \left(\lambda A_i + a \frac{\delta_i^2}{A_i}\right) \qquad (2)$$

is the discretized gravitational action and

$$E(l,s) = \frac{1}{2} \sum_{\langle ij \rangle} A_{ij} \left(\frac{s_i - s_j}{l_{ij}}\right)^2 \qquad (3)$$

is the energy of Ising spins $s_i = \pm 1$, which are located at the $N = L^2$ vertices $i$ of a triangulated torus with fixed coordination number $q = 6$. Here $\delta_i = 2\pi - \sum_{t \supset i} \theta_i(t)$ is the deficit angle and $A_i = \sum_{t \supset i} A_t/3$ is the baricentric area, where $\theta_i(t)$ denotes the dihedral angle at vertex $i$ and $A_t$ the area of a triangle $t$. Finally, $A_{ij} = \sum_{t \supset \langle ij \rangle} A_t/3$ is the area associated with a link $\langle ij \rangle$. In the first set of simulations we used the commonly employed simplest scale invariant functional measure $\mathcal{D}\mu(l) = \prod_{\langle ij \rangle} dl_{ij}/l_{ij}$. To investigate the dependence of our results on this choice we also performed exploratory simulations using the physically more motivated lattice version of the Misner measure $\mathcal{D}\mu(l) = \prod_{\langle ij \rangle} dl_{ij}/\sqrt{A_{ij}}$ [5]. Since both measures are scale invariant the total area is easily kept fixed at its initial value $A = \sum_i A_i = N$ by rescaling all link lengths when proposing a link update. In our simulations one MC step consisted of a single-hit Metropolis lattice sweep to update the link lengths $l_{ij}$ followed by a single-cluster flip to update (a fraction of) the spins $s_i$. In each run we recorded the energy density $e = E/A$ and the magnetization density $m = \sum_i A_i s_i/A$ in a time-series file. The statistical errors are computed by the jack-knife method.

Table 1
Critical parameters of the Ising transition.

| $a$ | $K_c$ | $U^*$ | $1/\nu$ | $\gamma/\nu$ | $\beta/\nu$ | $\alpha/\nu$ |
|---|---|---|---|---|---|---|
| 0.000 | 1.0234(2) | 0.609(33) | 0.95(2) | 1.717(6) | 0.123(4) | $-0.06(13)$ |
| 0.001 | 1.0265(1) | 0.612(5) | 1.00(1) | 1.735(5) | 0.127(3) | $-0.06(5)$ |
| 0.100 | 1.0295(1) | 0.615(6) | 0.98(1) | 1.728(3) | 0.123(2) | $-0.07(9)$ |

## 3. RESULTS

### 3.1. $dl/l$ measure

Here we concentrated on finite-size scaling (FSS) analyses for the $R^2$ couplings $a = 0$, 0.001, and 0.1, and performed long simulations with about 50 000 measurements near the critical coupling $K_c$ for various lattice sizes up to $L = 512$ [6]. We did not encounter any equilibration problems and the autocorrelation times of $e$ and $m$ turned out to be about $1-4$ measurements for all lattice sizes. By applying standard reweighting techniques we first determined the maxima of the susceptibility, $\chi = A(\langle m^2 \rangle - \langle |m| \rangle^2)$, the specific heat, $C = K^2 A(\langle e^2 \rangle - \langle e \rangle^2)$, as well as of $d\langle |m| \rangle/dK$, $d\ln\langle |m| \rangle/dK$, and $d\ln\langle m^2 \rangle/dK$. This defines five sequences of pseudo-transition points $K_{\max}(L)$ for which the scaling variable $x = (K_{\max}(L) - K_c)L^{1/\nu}$ should approach a constant for large $L$.

The critical exponent $\nu$ can then be estimated from (linear) least square fits to the FSS Ansatz $dU_L/dK \cong L^{1/\nu} f_0(x)$ or $d\ln\langle |m|^p \rangle/dK \cong L^{1/\nu} f_p(x)$ to the data at the various $K_{\max}(L)$, where $U_L = 1 - \langle m^4 \rangle/3\langle m^2 \rangle^2$. By averaging these estimates we obtain the values given in Table 1 which are all compatible with the Onsager value of $\nu = 1$. Assuming thus $\nu = 1$ we next determined $K_c$ from fits to $K_{\max}(L) = K_c + c/L$. Another, more precise method is to analyze the crossing points $K^\times$ of the curves $U_L(K)$ with $L$ and $L' = bL$. Combining this information we obtained the values quoted in Table 1 which were used in further analyses such as the asymptotic limit $U^*$ of $U_L(K_c)$. Within error bars our values in Table 1 agree with the very precise estimate for the regular square lattice, $U^* = 0.611(1)$ [9].

The exponent ratios $\gamma/\nu$ and $\beta/\nu$ follow from the FSS $\chi \cong L^{\gamma/\nu} f_3(x)$ and $\langle |m| \rangle \cong L^{-\beta/\nu} f_4(x)$ or $d\langle |m| \rangle/dK \cong L^{(1-\beta)/\nu} f_5(x)$, respectively. The final averages are again collected in Table 1. Also here we see little influence of the $R^2$ term. While our estimates for $\gamma/\nu$ are slightly below the Onsager value of 1.75, we obtain almost perfect agreement with the Onsager result $\beta/\nu = 0.125$. Finally we have checked that 3-parameter fits of the form $C_{\max} = A + BL^{\alpha/\nu}$ yield values of $\alpha/\nu$ consistent with zero; compare Table 1. Similar fits of $C$ at the other $K_{\max}(L)$ sequences as well as at $K_c$ gave compatible results.

Since the $R^2$ interaction term does not affect the critical exponents we give in Table 2 as final estimates the weighted average of the three simulations with different coupling $a$. For the $dl/l$ measure we definitely can exclude the DTS exponents and find strong evidence for the Onsager universality class. We conclude by mentioning that in the meantime we have also confirmed the failure of the Regge approach with $dl/l$ measure already for pure gravity [10].

Table 2
Critical exponents of the Ising transition.

|  | $\alpha$ | $\beta$ | $\gamma$ | $\nu$ |
|---|---|---|---|---|
| DTS | $-1$ | 0.5 | 2 | 1.5* |
| Onsager | 0 | 0.125 | 1.75 | 1 |
| Regge | $\approx 0$ | 0.126(2) | 1.75(2) | 1.01(1) |

Figure 1. Distribution of local deficit angles.

analyses of simulations with the Misner measure do not exclude the DTS exponents, we still have the hope that the Regge approach can survive as an alternative tool to describe quantum gravity.

## REFERENCES

1. V.A. Kazakov, JETP Lett. 44 (1986) 133, Phys. Lett. A119 (1986) 140 (1986); D.V. Boulatov and V.A. Kazakov, Phys. Lett. B186 (1987) 379.
2. J. Jurkiewicz, A. Krzywicki, B. Petersson, and B. Söderberg, Phys. Lett. B213 (1988) 511; S. Catterall, J. Kogut, and R. Renken, Phys. Rev. D45 (1992) 2957; C. Baillie and D. Johnston, Mod. Phys. Lett. A7 (1992) 1519; R. Ben-Av, J. Kinar, and S. Solomon, Int. J. Mod. Phys. C3 (1992) 279.
3. V.G. Knizhnik, A.M. Polyakov, and A.B. Zamalodchikov, Mod. Phys. Lett. A3 (1988) 819.
4. T. Regge, Nuovo Cimento 19 (1961) 558.
5. M. Gross and H. Hamber, Nucl. Phys. B364 (1991) 703.
6. C. Holm and W. Janke, Phys. Lett. B335 (1994) 143.
7. W. Bock and J.C. Vink, preprint UCSD 94-8; and this volume.
8. H. Hamber, Les Houches 1984, ed. K. Osterwalder and R. Stora (North Holland, Amsterdam, 1986); p. 375.
9. D.W. Heermann and A.N. Burkitt, Physica A162 (1990) 210.
10. C. Holm and W. Janke, this volume.